\documentclass[a4paper,11pt]{article}
\usepackage[margin=2.5cm]{geometry}
\usepackage[utf8]{inputenc}
\usepackage{amssymb,amsmath,epsfig,esint}
\usepackage{CJKutf8}
\usepackage{amsfonts}
\usepackage{setspace,lipsum}
\usepackage{adjustbox}
\usepackage{graphicx}
\usepackage{subcaption}
\usepackage[english]{babel}
\usepackage{booktabs}
\usepackage{cellspace}
\usepackage{multicol}
\usepackage{multirow}
\usepackage{float,tabularx,array}
\usepackage{authblk}
\usepackage{fourier} 
\usepackage{array}
\usepackage{makecell}
\usepackage[nottoc]{tocbibind}
\usepackage[colorlinks]{hyperref}
\usepackage{url}
\hypersetup{
    colorlinks=true,
    linkcolor=red,
    filecolor=magenta, 
    citecolor=blue,        
    urlcolor=blue,
    pdftitle={Coupled scalar field cosmology with effects of curvature },
    pdfpagemode=FullScreen,
    }

\usepackage[noadjust]{cite}

\newcolumntype{C}[1]{>{\centering\arraybackslash}p{#1}}

\begin{document}
\title{\bf Coupled scalar field cosmology with\\ effects of curvature }
\author{Trupti Patil\thanks{trupti19@iiserb.ac.in} \hspace{0.1cm} and \hspace{0.1cm}Sukanta Panda
\thanks{sukanta@iiserb.ac.in} \\
$^\ast$$^\dagger$Department of Physics, IISER Bhopal,\\
Bhopal - 462066, India\\
}

\date{}
\maketitle
\begin{abstract}
In the present study, we investigate the interaction between dark energy and dark matter, particularly emphasizing the effects of curvature in the realm of  Friedmann-Lema$\hat{\imath}$tre-Robertson-Walker (FLRW) space-time. We examine the system by forming a dynamic set of equations for various critical points. Later, we study their stability characteristics and show that with a suitable choice of potential, the system gives rise to the late-time attractor (stable) solution in the expanding environment. Lastly, we present the cosmological compatibility of the model using the phase-space portrait tools.
\end{abstract}

\section{Introduction} \label{1}
The cosmological and observational evidence that the universe is currently under an accelerated expansion has been put forward in various notable works \cite{Riess_1998,Dunkley_2009,Boughn:2004zm,PhysRevD.69.103501,2007,DiValentino:2019jae,2006}. In the literature, the cosmologists investigated ample theoretical models to support such observational analysis concerning modern cosmology \cite{article,Wang_2016,Perivolaropoulos:2021jda}. The mystery behind the unknown fundamental nature of the dark sector entities inspired many researchers to develop dark sector models in association with cosmological observations \cite{Gavela:2009cy,Yang:2018xah,Johnson:2021wou,10.1093/mnras/stx2278,10.1093/mnras/stv1495,salv,Olivares:2007rt}. The cosmological frameworks; that signifies the exchange of energy between the dark matter and dark energy are considered useful to alleviate both, the cosmological constant problem and coincidence problem \cite{PhysRevLett.82.896,2020,RevModPhys.61.1,Doran:2002ec,2009,2016,2014} and thus, makes it an interesting study to probe further. Several phenomenological fluid models \cite{Wang_2016,Marra:2015iwa,Odintsov:2018uaw,Oikonomou:2019boy,Yang:2021oxc,Odintsov:2017icc,Pal:2019bsm} and scalar fields models \cite{PhysRevD.104.063517,potting2021coupled,GonzlezDaz2000CosmologicalMF,refId0,PhysRevD.63.103510,Bamba:2012cp,Harko2014ArbitrarySA,Oikonomou:2019muq,ROZASFERNANDEZ2012313} have extensively been studied in this realm. Modified gravity (See Ref.\cite{Rudra:2014xba,Vinutha:2021jbs,Clifton:2011jh} and references therein) is another way to address such issues.
\par
Inspired by the above profound groundwork, we have constructed an interacting dark energy-dark matter model based upon the theory of chiral cosmology \cite{Chervon2013ChiralCM,Paliathanasis:2018vru,Paliathanasis_2020} in the earlier work \cite{Patil:2022uco}. We have studied the formation of an autonomous system from the coupled non-linear differential equations for the dark energy-dark matter interaction (DDI) model through field-fluid analogy (see the references therein \cite{PhysRevD.103.123517,PhysRevD.103.023510}). We found the physically acceptable equilibrium points and examined their stability in detail. Our findings suggested that the stable attractor solution leading to an accelerated expanding universe with the effective equation of state (EoS) $``\omega_{eff}$" less than $-1$ corresponding to phantom behaviour is also achievable up until the background dynamics are studied. It perfectly pictures a sufficiently long, extended era of the matter-dominated universe following the advancement of an accelerated expansion in the present time. As we move forward, we also notice that the impact of curvature term on the cosmological dynamics cannot be underrated. Thus, to understand the effects of the curvature, we now examine the system with the coupling between two scalar fields that represent the dark sector of the universe in the case of the FLRW (homogeneous and isotropic) universe with non-zero curvature. We also form a one-to-one analogy between the field theory method and the fluid approach of the interacting dark sector. The analogy assists us in achieving a unique interaction function $`Q$' derived 
rigorously from the series of equations starting from action integral to continuity equations where one of the fields becomes a dark matter substitute. The mathematical background behind the non-linear coupled differential equations transpiring during the analysis is an impressive technique for understanding the qualitative features of the cosmological model under analysis.
We obtain these equations by defining preferable dimensionless variables, and examination of these is essential to understand the mystifying dynamics of the observable universe in depth. As discussed, many such theoretical models both in flat space-time \cite{BAHAMONDE20181,Boehmer:2008av,Caldera-Cabral:2008yyo,Chakraborty:2020vkp,Chakraborty:2020yfe,2021,Singh:2019enu,Mandal:2021ekc,Chakraborty:2021pkp,Lazkoz:2006pa} and curved space-time \cite{Pavlov:2013nra,Tot:2022dpr,Paliathanasis:2022luh} have been built and examined by the cosmologists to reveal more about the dark sector phenomenon. In \cite{Pavlov:2013nra}, they have studied the non-flat time-variable DE model to constrain space curvature using various cosmological observables. Curved cosmology in \cite{Tot:2022dpr} studies the curvature effects on the inflationary
properties and on the bouncing solutions of the model. The Non-zero curvature model studied in \cite{Paliathanasis:2022luh} aims at solving the flatness problem by showing the stable hyperbolic inflationary solutions. On the other hand, in our work, we investigate the non-zero curvature model to study the late-time attractor solutions in the accelerated expanding regime.
\par
The motivation behind the proposed work is the idea of comprehending whether introducing a curvature term in the dynamical system would affect the local stability and late-time cosmological behaviour during the background study of the model. We found new dynamics due to the presence of curvature term when compared to the flat case DDI model, as this modifies the obtained critical points. 
These critical points have subsequently been discussed in the manuscript. The study also supports the physical interpretation of the model via the geometrical display of the evolution of various key cosmological parameters. \\
We organize the paper as follows: In the next section, we set up and review the coupled dark matter (DM) and dark energy (DE) scalar field scenario. We also set up a framework for the one-to-one correspondence between the field-fluid approaches. Section \ref{3} describes the dynamical analysis approach used to investigate the attractors in the expanding regime. In the subsection, we present the parameter constraints on the interacting model and discuss the physical results concerning the background cosmology. Finally, we conclude the work with a brief discussion in Section \ref{4}.
\par

\section{Coupled dark sector model} \label{2}
The action integral for the dark sector coupling is motivated from \cite{Patil:2022uco}.
\begin{equation}
\label{eqn:action integral}
    S =  \int d^4x \sqrt{-g}\bigg( \frac{R}{2} + \frac{1}{2} g^{\mu \nu} \nabla_{\mu}\phi \nabla_{\nu}\phi - \frac{1}{2} B(\phi) g^{\mu \nu} \nabla_{\mu}\psi \nabla_{\nu}\psi  - V(\phi) - \alpha \phi^2 \psi^2\bigg).
\end{equation}
In order to achieve our goal of designing a dark sector interaction model, we consider one of the fields as a dark matter candidate and another as a dark energy candidate. The details about the roles played by the fields will naturally come across as we progress further. \\

Following the analysis, we write the FLRW space-time for the background space as:
\begin{equation}
\label{eqn:line element}
    ds^2 = -dt^2 + a^2(t) \Big[\frac{dr^2}{1-\epsilon r^2} + r^2 d\theta^2 + r^2 sin^2\theta d\phi^2  \Big].
\end{equation}
Where $\epsilon$ is the Gaussian curvature of the space at the time when a(t)=1 and a(t) is the scale factor as a function of cosmic time. When $\epsilon$ is zero, the universe is flat, whereas non-zero $\epsilon$ produces curved space-time. We have a closed (spherical) universe for positive curvature and an open (hyperbolic) universe for negative curvature. The previous study \cite{Patil:2022uco} analyzed the action integral (\ref{eqn:action integral}) for $\epsilon = 0$ case. Unlike the earlier analysis, here we investigate the non-zero curvature case.\\
The variation of action (Eq.\ref{eqn:action integral}) w.r.t. the inverse metric brings us Einstein’s gravitational field equations as:
\begin{multline}
\label{eqn:gravity field eq}
    G_{\mu \nu} = T_{\mu \nu} \equiv \bigg(-\nabla_{\mu} \phi \nabla_{\nu} \phi + \frac{1}{2} g_{\mu \nu} \nabla_\xi \phi \nabla^\xi \phi - g_{\mu \nu} V(\phi) \\
    +B(\phi) \nabla_{\mu} \psi \nabla_{\nu} \psi  - \frac{1}{2} B(\phi) g_{\mu \nu} \nabla_\xi \psi \nabla^\xi \psi - g_{\mu \nu} \alpha \phi^2 \psi^2 \bigg)
\end{multline}
where $T_{\mu \nu}$ is the total energy-momentum tensor.\\

Also, from the line element (\ref{eqn:line element}) and the action integral (\ref{eqn:action integral}), the Friedmann equations and the conservation equations for non-zero curvature are given as:
\begin{equation}
\label{eqn:Friedmann eq1}
\begin{split}
\frac{-\dot{\phi}^2}{2} + \frac{B(\phi)}{2}\dot{\psi}^2 + V(\phi) + \alpha \phi^2 \psi^2 -\frac{3 \epsilon}{a^2} = 3H^2,  \\[4pt]
-\bigg( \frac{-\dot{\phi}^2}{2} + \frac{B(\phi)}{2}\dot{\psi}^2 - V(\phi) - \alpha \phi^2 \psi^2  \bigg) -\frac{ \epsilon}{a^2} = 2\dot{H} + 3H^2.  \\[05pt]
\end{split}
\end{equation}
\begin{equation}
\label{eqn:continuity eq1}
\begin{split}
-\ddot{\phi}-3H\dot{\phi}-\frac{B,_\phi}{2} \dot{\psi}^2 + 2 \alpha \phi \psi^2 + V,_\phi = 0,  \\[4pt]
\ddot{\psi}+3H\dot{\psi}+\frac{B,_\phi}{B} \dot{\phi} \dot{\psi} + 2 \alpha \psi \phi^2 B^{-2} = 0. \\[05pt]
\end{split}
\end{equation}
where, $H=\frac{\dot{a}}{a}$ is the Hubble rate.\\
Given the coupling between these fields, it is found that the dark energy (DE) and dark matter (DM) do not satisfy the energy-momentum tensor conservation equation independently; instead, they satisfy the local conservation equation in the form given as: 
\begin{equation}
\label{eqn: local continuity eq}
    -\nabla^\mu T_{\mu \nu} ^{(\phi)} = Q_\nu= \nabla^\mu T_{\mu \nu} ^{(\psi)},
\end{equation}
where,
\begin{equation}
\label{eqn: interaction eq1}
    Q_\nu= \nabla^\mu T_{\mu \nu} ^{(\psi)} = -\bigg(\frac{B,_\phi}{2} \nabla_\xi \psi \nabla^\xi \psi + 2 \alpha \phi \psi^2 \bigg) \nabla_\nu \phi,
\end{equation}
characterizes the energy shifting between dark energy and dark matter in interacting dark zone. Here, $T_{\mu \nu} ^{(\psi)}$ and $T_{\mu \nu} ^{(\phi)}$ play the role of energy-momentum tensor of field $\psi$ and field $\phi$, respectively.\\
\subsection{Fluid representation of the \textbf{DDI} model}
The field theory application is likely the fundamental description of the system. However, the fluid description is more advantageous to study the observational field. In that regard, we develop one-to-one correspondence between the field theory approach and the fluid approach \cite{PhysRevD.103.023510} of the interacting dark sector. With this, we try to illustrate the unique form of the interaction function $`Q_\nu$' systematically obtained by using an entire set of equations from field action to conservation equations where one of the fields becomes dark matter representative.
\par
Considering such a description of the above coupled DE-DM model, it is often suitable to frame the dark matter as a fluid component. The work openness provides the freedom to assume either the field $``\phi$'' or the field $``\psi$'' as a dark matter entity. Therefore, we now replace the dark matter fluid for the scalar field $``\psi$''. This exchange sets the field $``\phi$'' as a DE component. Following the analysis, the dark matter energy density $\rho_{dm}$ and pressure $P_{dm}$ develop as
\begin{eqnarray}
\rho_{dm}=-\frac{1}{2} B(\phi) \bigg(g^{\mu \nu} 
\nabla_{\mu} \psi \nabla_{\nu} \psi - 2 \alpha \phi^2 \psi^2 B^{-1} \bigg), \label{eq:19}\\
P_{dm}=-\frac{1}{2} B(\phi) \bigg(g^{\mu \nu} 
\nabla_{\mu} \psi \nabla_{\nu} \psi + 2 \alpha \phi^2 \psi^2 B^{-1} \bigg). \label{eq:20}
\end{eqnarray}
Whereas the energy-momentum tensor \eqref{eqn:gravity field eq} for the DDI model in terms of DE scalar field ($\phi$) and DM fluid can be revised as:
\begin{equation}
\label{eqn:21}
    T_{\mu \nu}= \bigg(-\nabla_{\mu} \phi \nabla_{\nu} \phi + \frac{1}{2} g_{\mu \nu} \nabla_\xi \phi \nabla^\xi \phi - g_{\mu \nu} V(\phi) + P_{dm}  g_{\mu \nu} + ( \rho_{dm} + P_{dm})  u_{\mu}  u_{\nu}\bigg),
\end{equation}
where $u_{\mu}$ is termed as the four-velocity of the DM fluid.
\par
Theoretical findings show that the interaction rate is normally proportional to the dark energy density, dark matter density, or both in the phenomenological fluid models. Apart from this, the interaction rate $``Q_\nu$'' is set manually. Unlike in fluid models, the interaction rate or the interaction term $``Q_\nu$'' derived in \eqref{eqn:interaction eq2} has a particular form obtained by correlating the fields and fluids for the present dark sector coupling model. Taking this into account, $`` Q_\nu$'' in Eq.\eqref{eqn: interaction eq1} is now transformed as:
\begin{equation}
\label{eqn:interaction eq2}
    Q_\nu = \bigg( \frac{B,_{\phi}}{B}  -  \frac{2}{\phi} \bigg)\frac{\rho_{dm}}{2} \nabla_{\nu} \phi,
\end{equation}

The above equations, Eq.\eqref{eqn:Friedmann eq1} $\&$ Eq.\eqref{eqn:continuity eq1} are also transformed as:
\begin{equation}
\label{eqn:Friedmann eq2}
\begin{split}
\rho_{dm} - \frac{\dot{\phi}^2}{2}+ V(\phi) -\frac{3 \epsilon}{a^2} = 3H^2, \\[4pt]
-\bigg(P_{dm}- \frac{\dot{\phi}^2}{2}- V(\phi) \bigg) -\frac{ \epsilon}{a^2} = 2\dot{H} + 3H^2. \\[05pt]
\end{split}
\end{equation}
\begin{equation}
\label{eqn:continuity eq2}
\begin{split}
-\ddot{\phi} \dot{\phi}-3H\dot{\phi}^2+V,_\phi \dot{\phi}= Q,\\[4pt]
\dot{\rho}_{dm}+3H\rho_{dm}=-Q. \\[05pt]
\end{split}
\end{equation}

Here, we note that field $``\phi$'', the DE component possesses negative kinetic energy that can realize the equation of state parameter of dark energy $\omega_\phi < -1$ in their evolution. This feature exhibits phantom field behaviour. It is correct that the field theory with negative kinetic term questions the widely accepted energy conditions and leads to rapid vacuum decay \cite{PhysRevD.68.023509}, but it is still interesting to study these models as they are phenomenologically interesting. Also, models with such an exotic form of energy with $\omega_\phi < -1$ are very much acknowledged by the observations \cite{Hannestad:2002ur,Melchiorri:2002ux,Lima:2003dd,Alam:2004jy,Alam:2003fg,Wang:2003gz} and leads to the required late time expanding universe. Thus, the motivation to examine such a cosmological coupling model is strongly driven by the observations.

\section{Dynamical system approach}\label{3}
We formulate an autonomous set of equations to study the qualitative dynamics of cosmological evolution. Accordingly, we introduce the following dimensionless variables
\begin{align}
\label{eqn:10}
    x &=\frac{\dot{\phi}}{\sqrt{6}H},&
    y &=\frac{{\sqrt{V(\phi)}}}{\sqrt{3}H},&
    \Omega_{m} &=\frac{\rho_{dm}}{3H^2},&
    \Omega_{k} &= -\frac{\epsilon}{a^2 H^2},
\end{align}
satisfying the constraint equation
\begin{equation}
\label{eqn:11}
    -x^2 + y^2 + \Omega_{m} + \Omega_{k} = 1 .
\end{equation}
Using Eq. \eqref{eqn:10}, we write the cosmological parameters as expressed below\\
DE density parameter: 
\begin{equation}
\label{eqn:12}
    \Omega_\phi = -x^2+y^2.
\end{equation}
DE equation of state (EoS) parameter $\&$ effective EoS parameter :
\begin{equation}
\label{eqn:eosphi}
    \omega_\phi = \frac{-x^2-y^2}{-x^2 +y^2},
\end{equation}
\begin{center}
    {\&}
\end{center}
\begin{equation}
\label{eqn:eoseffective}
    \omega_{eff} = \frac{1}{3} \big(-1-4x^2 -2y^2 +\Omega_{m} \big).
\end{equation}
Operating derivative with respect to number of e-foldings $N = ln(a)$ on above variables (Eq.\ref{eqn:10}), we can derive the following autonomous system. During the dynamical analysis, we assumed an exponential potential, $V(\phi) \propto e^{\alpha (\phi)}$ that makes the equation corresponding to $\lambda$, $\lambda'$ disappear. Similarly, assumption of the coupling term $B(\phi)$ proportional to $ e^{\lambda (\phi)}$ and of non-zero constant `$ \alpha (\phi)$' results in the 4D complete autonomous system as:
\begin{equation}
\label{eqn:autonomous eqns}
\begin{split}
x^\prime= - \sqrt{\frac{3}{2}}\bigg(\lambda y^2 +\frac{I}{x}\bigg)-\frac{3}{2}x \bigg(1+x^2+y^2 + \frac{\Omega_{k}}{3} \bigg), \\
y^\prime= -\sqrt{\frac{3}{2}}(\lambda x y) +\frac{3}{2}y \bigg(1-x^2-y^2 -\frac{\Omega_{k}}{3} \bigg), \\
\Omega_{m}^\prime=  -\sqrt{6}I -3 \Omega_{m} \bigg(x^2 + y^2 + \frac{\Omega_{k}}{3} \bigg),  \\
\Omega_{k}{^\prime}= \Omega_{k} (1- 3x^2 -3y^2 -\Omega_{k}),\\
\end{split}
\end{equation}
Where $`I$' represents the scaled interaction term
\begin{equation}
\label{eqn:16}
    \textit{I}=\frac{(2 \beta - k)}{2} x \cdot \Omega_{m} = \frac{1}{3}\frac{Q}{\sqrt{6}H^3},
\end{equation}
with $\lambda$, k, $\beta$ and $\alpha$ are defined as follows
\begin{equation*}
 \label{eqn:17}
     \lambda(\phi)=-\frac{V,_\phi}{V(\phi)}, \qquad k(\lambda) =-\frac{B,_\phi}{B(\phi)}, \qquad \beta(\phi) =-\frac{\alpha,_\phi}{\alpha(\phi)}, \qquad \alpha = \alpha(\phi).
\end{equation*}
%
%
%
\subsection{Critical points and phase space analysis}\label{3.1}
The equilibrium points for the reduced 4D dynamical system are manifested in Table \ref{tab:critical points table}.\\
We examine each equilibrium point and discuss its stability and other physical attributes in the following manner.
\par
\begin{figure}[!ht]
  \centering
  \begin{subfigure}[b]{0.42\linewidth}
    \includegraphics[width=\linewidth]{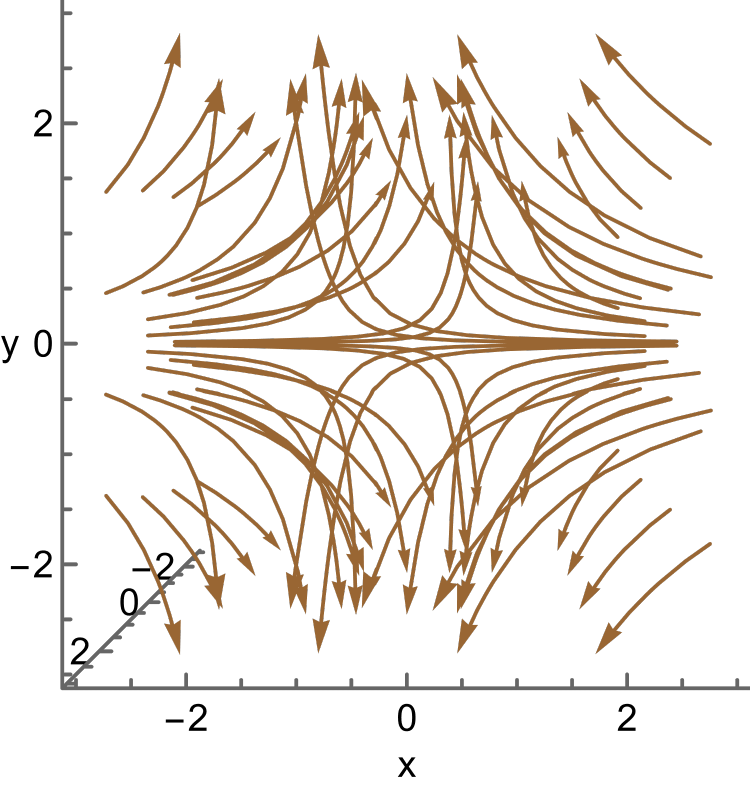}
     \caption{x-y-$\Omega_{m}$ -plane}
     \vspace{1cm}
     \label{fig:plotxyz}
  \end{subfigure}
  \begin{subfigure}[b]{0.42\linewidth}
    \includegraphics[width=\linewidth]{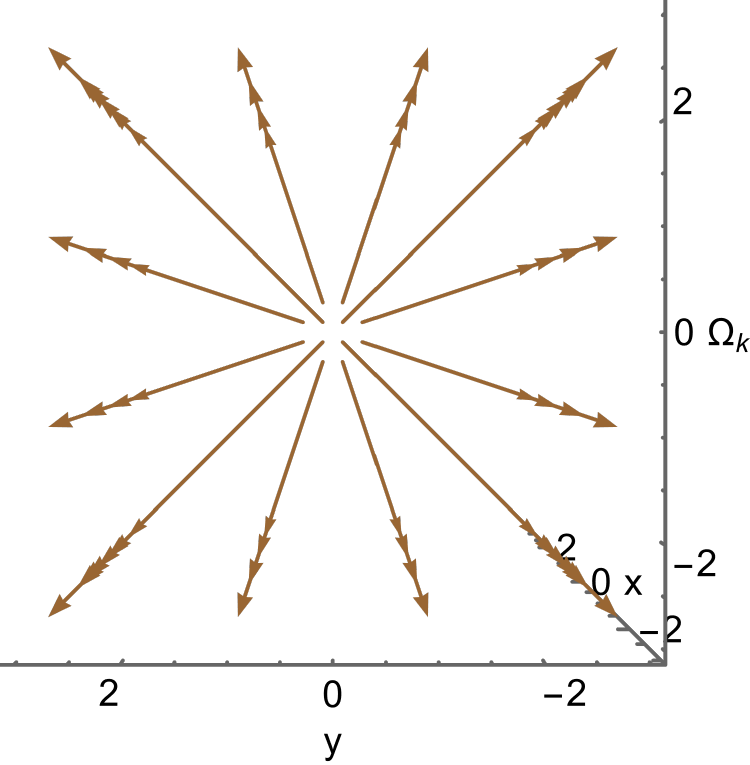}
    \caption{{x-y-$\Omega_{k}$}-plane}
    \vspace{1cm}
    \label{fig:plotxyomegak}
  \end{subfigure}
  \begin{subfigure}[b]{0.44\linewidth}
    \includegraphics[width=\linewidth]{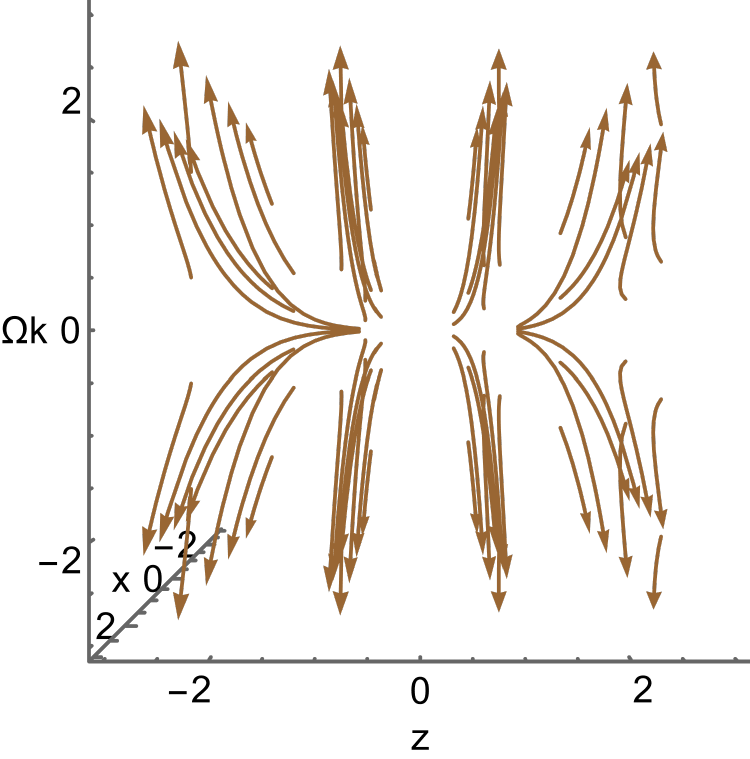}
    \caption{{x-$\Omega_{m}$-$\Omega_{k}$}-plane}
    \label{fig:plotxzomegak}
  \end{subfigure}
  \begin{subfigure}[b]{0.44\linewidth}
    \includegraphics[width=\linewidth]{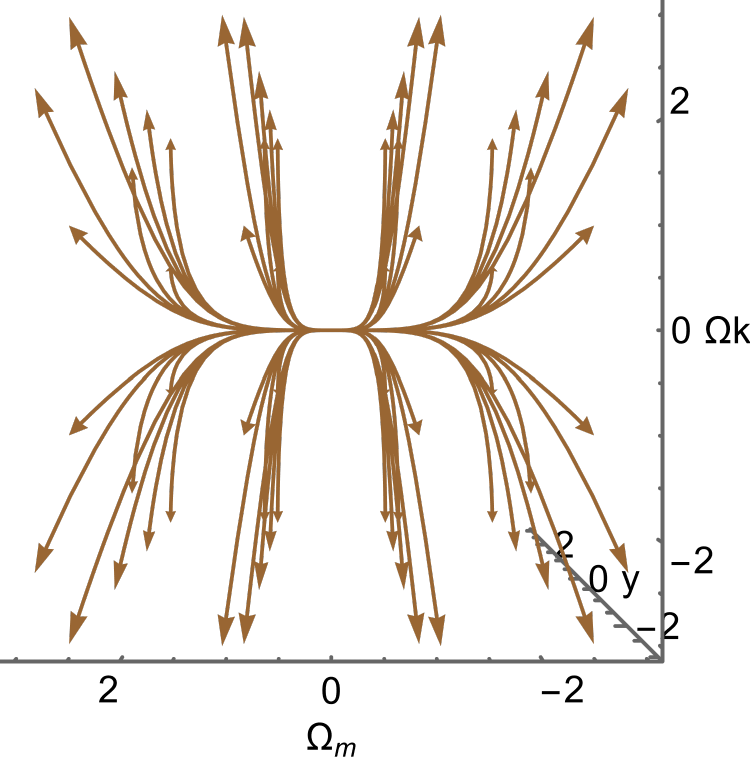}
    \caption{{y-$\Omega_{m}$-$\Omega_{k}$}-plane}
    \label{fig:plotyzomegak}
  \end{subfigure}  
  \caption{Three-dimensional phase portraits corresponding to critical point A when $\beta=k=1$. \textbf{(a)} portrays projection of vector fields on \textbf{x-y-\boldmath{$\Omega_{m}$}}-plane (saddle-node), \textbf{(b)} portrays projection of vector fields on \textbf{x-y-\boldmath{$\Omega_{k}$}}-plane (unstable node), \textbf{(c)}  portrays projection of vector fields on \textbf{x-\boldmath{$\Omega_{m}$}-\boldmath{$\Omega_{k}$}}-plane (unstable node)  and \textbf{(d)} portrays projection of vector fields on \textbf{y-\boldmath{$\Omega_{m}$}-\boldmath{$\Omega_{k}$}}-plane (unstable node).  Arrows represent the direction of the flow along the trajectories.}
  \label{fig:pointA_phaseportrait}
\end{figure}

\textbf{Point A} is of hyperbolic nature with eigenvalues $\Big(\frac{5}{3},\hspace{0.1cm} \frac{5}{3},\hspace{0.1cm} \frac{-4}{3}, \hspace{0.1cm} \frac{1}{3} \Big)$. The point is a saddle-node, attracting nearby trajectories in some directions and repelling them along the others. The condition, $\omega_{eff}=-\frac{1}{3}$ implies the solution that describes the Milne-like universe. We display the flow of the vector fields on 3D phase-space in all possible respective coordinate systems in Fig.\ref{fig:pointA_phaseportrait}. From Table \ref{tab:critical points table}, it is found that point A is representative of the curvature-dominated epoch. Nonetheless, this point's saddle nature determines that the curvature domination period is simply a transitory period in the history of cosmology.
\begingroup
\begin{table}[!ht]
\begin{center}
\begin{adjustbox}{width=1\textwidth}
\small
    \begin{tabular}{c| c c c c c c}
      \hline
      \toprule
      \thead{Critical\\ Points} &  \thead{x}  &  \thead{y} & \boldmath{$\Omega_{m}$} &  \boldmath{$\Omega_{k}$}  & \boldmath{$\omega_{eff}$}\\
      \bottomrule
      \hline
      A &  \makecell{0} &  \makecell{0} &  \makecell{0} &  \makecell{1} &  \makecell{$-\frac{1}{3}$} \\
      B &  \makecell{$-\frac{\sqrt{\frac{2}{3}}}{(2\beta-k)}$} &  \makecell{0} &  \makecell{$\frac{8}{3(2\beta-k)^2}$} &  \makecell{$1-\frac{2}{(2\beta-k)^2}$} &  \makecell{$-\frac{1}{3}$} \\
      C &  \makecell{$-\frac{(2\beta-k)}{\sqrt{6}}$} &  \makecell{0} &  \makecell{$1+\frac{(2\beta-k)^2}{6}$} &  \makecell{0} &  \makecell{$-\frac{1}{6}(2\beta-k)^2$} \\
      D &  \makecell{$\frac{-\sqrt{6}}{(2\beta-k-2\lambda)}$}  &  \makecell{$\sqrt{\frac{-6}{(2\beta-k-2\lambda)^2}+\frac{(2\beta-k)}{(2\beta-k-2\lambda)}}$} &  \makecell{$\frac{12}{(2\beta-k-2\lambda)^2}-\frac{2\lambda}{(2\beta-k-2\lambda)}$} &  \makecell{0} &  \makecell{$-\frac{(2\beta-k)}{(2\beta-k-2\lambda)}$}\\
      E & \makecell{$-\frac{\lambda}{\sqrt{6}}$} &  \makecell{$\sqrt{1+\frac{\lambda^2}{6}}$} &  \makecell{0} &  \makecell{0} &  \makecell{$-1-\frac{\lambda^2}{3}$} \\  \bottomrule \hline
    \end{tabular}
    \end{adjustbox}
  \end{center}
  \caption {\label{tab:critical points table}Critical points and cosmological parameters relative to the reduced 4D autonomous system for a given $\lambda$ and coupling function $B(\phi)$.}
  \end{table}
\endgroup
%
%
\par
\begin{figure}
  \centering
  \begin{subfigure}[b]{0.48\linewidth}
    \includegraphics[width=\linewidth]{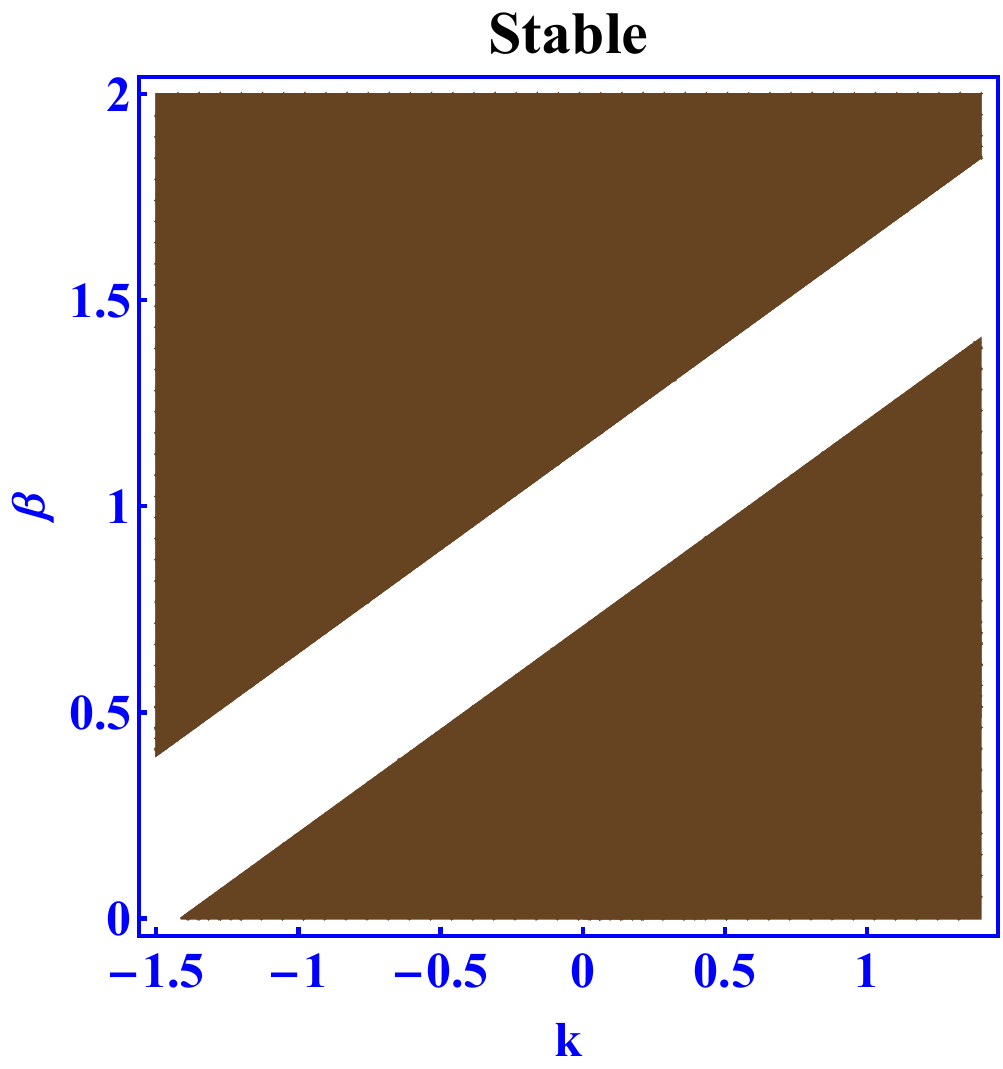}
     \caption{}
     \label{fig:pointB_stable}
  \end{subfigure}
  \begin{subfigure}[b]{0.482\linewidth}
    \includegraphics[width=\linewidth]{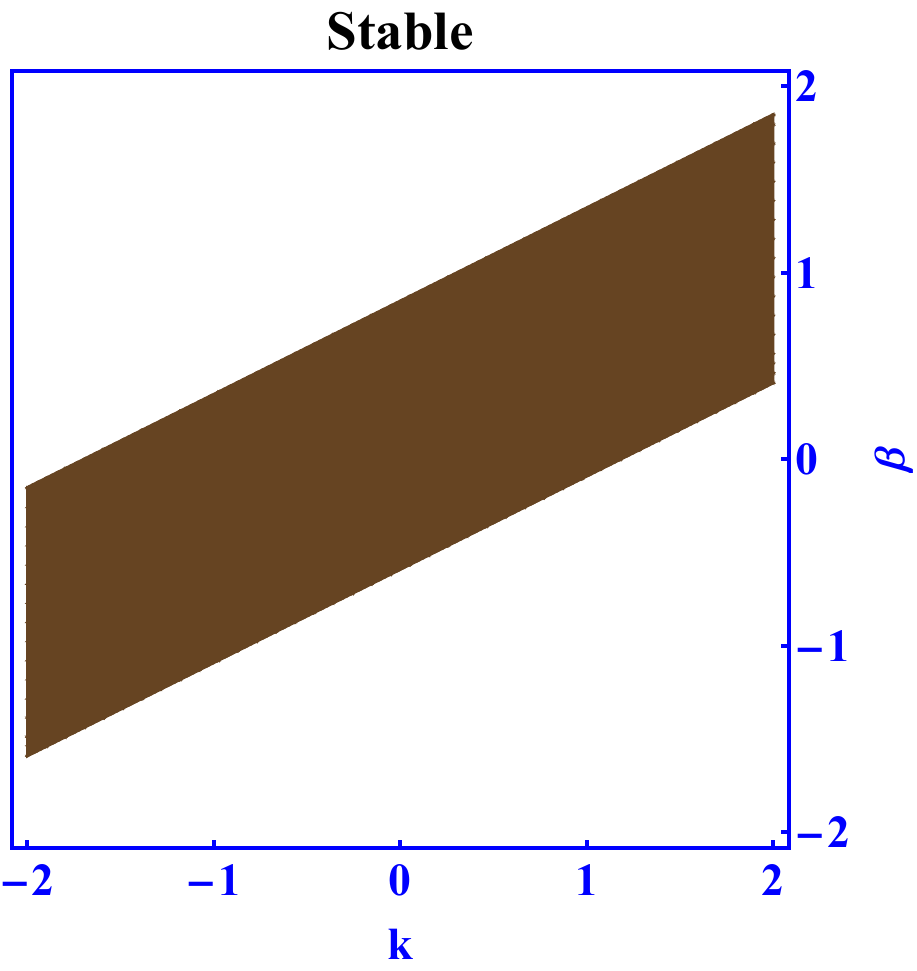}
    \caption{}
    \label{fig:pointD_stable}
  \end{subfigure}
  \caption{The brown shaded region of the parameter space ($\beta, k$) shows the existence of stable features (a) for equilibrium point B and (b) for equilibrium point D. The phase plane is plotted with $\lambda=0.4$.  }
    \label{fig:stability plots}
\end{figure}

\textbf{Point B} represents a hyperbolic point. The behaviour of this point is the same as the fixed point A with $\omega_{eff}=-\frac{1}{3}$. However, the eigenvalues (See \ref{Appendix}), in this case, are unsuitable for further analysis restricting to determine the nature of the point accurately. Nevertheless, for a comprehensive analysis, we follow an analytical approach to show the stable properties of this point. In Fig. \ref{fig:pointB_stable}, we showcase the region of parameter space ($\beta, k$) where point B exhibits the stability feature. 
\par
\textbf{Point C} is real and physically acceptable for any real value of parameters $\beta$ and $k$. In the parameter space constraint, $\sqrt{2}<(2\beta-k)<-\sqrt{2}$ with $\beta = k\ne 0 $, the point describes an accelerated solution of the expanding universe. The eigenvalues corresponding to the Jacobian matrix of the fixed point C are:
\begin{equation*}
    \Bigg( -\frac{1}{2}(2\beta-k)^2, \quad 1-\frac{1}{2}(2\beta-k)^2, \quad -\frac{3}{2}-\frac{1}{4} (2\beta-k)^2, \quad \frac{3}{2}-\frac{1}{4}(2\beta-k)(2\beta-k-2\lambda) \Bigg).
\end{equation*}
Point C is a hyperbolic equilibrium point. It characterizes stable behaviour when $  (2\beta-k - 2\lambda) \geq 3\sqrt{2}$ \hspace{0.1cm} and \hspace{0.1cm} $\bigg( (2\beta-k)>\sqrt{2} $ \hspace{0.1cm} or\hspace{0.1cm} $  (2\beta-k) < -\sqrt{2} \bigg)$. Similarly it characterizes saddle behaviour when $ (2\beta-k - 2\lambda) \leq 3\sqrt{2}$ \hspace{0.1cm} and\hspace{0.1cm} $ \bigg((2\beta-k)<\sqrt{2} $ \hspace{0.1cm}or\hspace{0.1cm} $(2\beta-k)> -\sqrt{2} \bigg) $.
\par
\textbf{Point D} is valid for $-2 < \frac{2\lambda}{(2\beta-k)}<1$ in the phase-space and substantiates the possibility of the accelerated universe within the limit as mentioned earlier for $\frac{2\lambda}{(2\beta-k)}$. The point gives a constant ratio between DE and DM density parameters describing the scaling solution in the phase space.
\begin{equation*}
         \frac{\Omega_{\phi}}{\Omega_m} = \frac{-3+ \beta^2-  \beta \lambda}{3+\lambda^2-  \beta \lambda}.
\end{equation*}\\[1pt]
For $0<\frac{2\lambda}{(2\beta-k)}<1$, point D features a phantom field dominated universe with $\omega_{eff}<-1$. On the other hand, the constraint, $-2<\frac{2\lambda}{(2\beta-k)}<0$ describes a quintessence dominated universe with $-1<\omega_{eff}<-\frac{1}{3}$.
\vspace{0.1cm}
\newline
Accordingly, when $\frac{2\lambda}{(2\beta-k)} \rightarrow 0$, point D describes the solution where the universe evolves like a de Sitter universe dominated by a cosmological constant with $\omega_{eff} =-1$.\\
Moving along the stability note, as explained in the above case for point B, we found that the complexity of the eigenvalues corresponding to the Jacobian matrix of point D (See \ref{Appendix}) obstructs us from carrying out further analysis, required to decide the nature of the fixed point. Therefore, here as well, with a comprehensive purpose, we show the region on the parameter space ($\beta, k$) (Fig. \ref{fig:pointD_stable}) where equilibrium point D is an attractor (stable) point.
\par
\begin{figure}[ht!]
    \centering
    \includegraphics[width=0.6\textwidth]{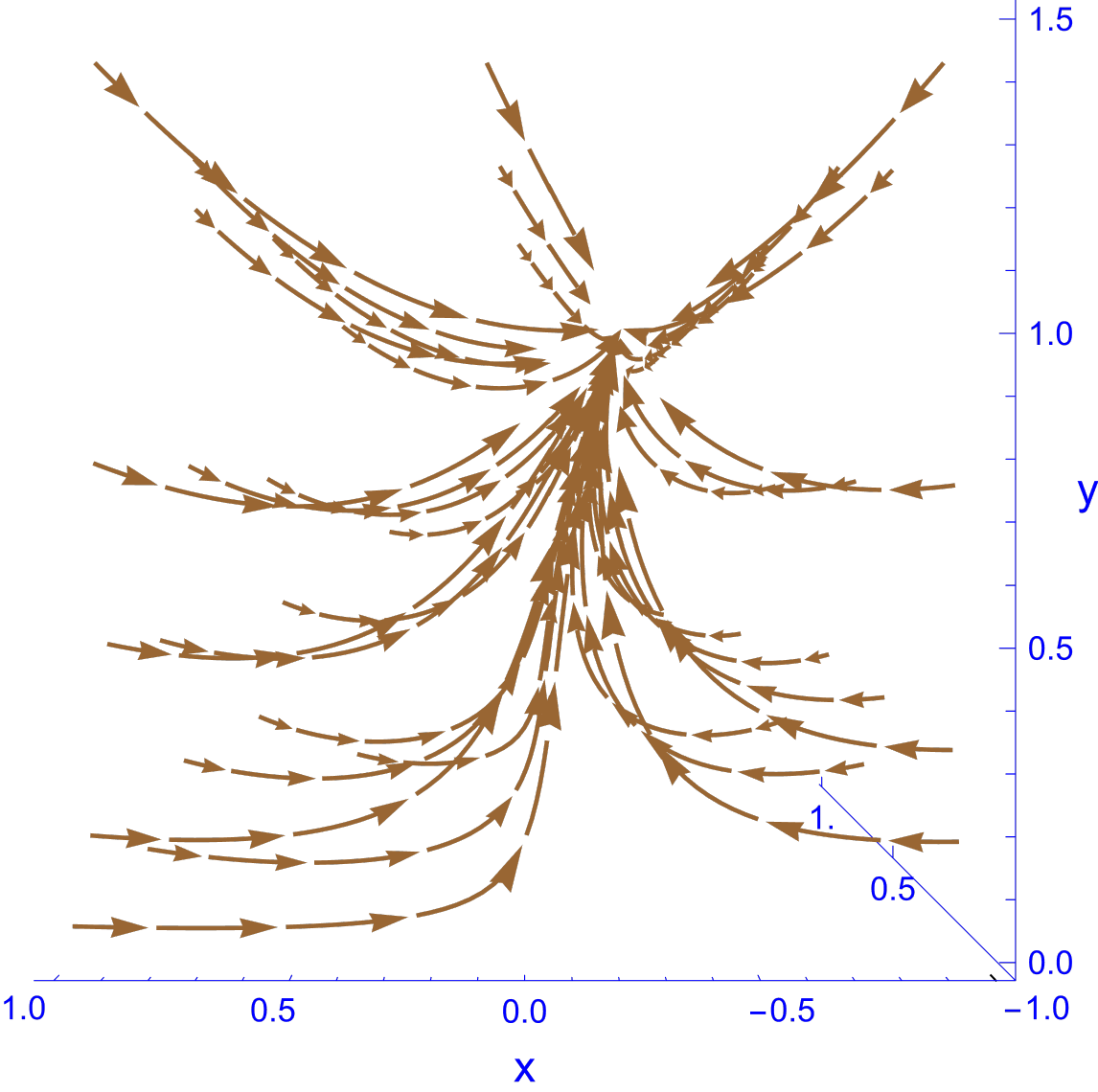}
    \caption{Three-dimensional phase space portrait of the dynamical system in the plane (x,y,$\Omega_k$) for an equilibrium point E when $\lambda=0.4$ . The third axis represents $\Omega_k$.  }
    \label{fig:3D_attractorE}
\end{figure}

\textbf{Point E} is physically acceptable for all real values of $\lambda$ and describes an accelerating solution for parameter constraint $\lambda^2 >-2$. It also pictures the absolute dominance by scalar field $`\phi$'. The point exhibits various (quintessence, phantom, cosmological constant) field dominated behaviour. For $\lambda^2 >0$ and $-2<\lambda^2<0$, we achieve a phantom field and quintessence field dominated universe while in the limit $\lambda \rightarrow 0$, scalar field behaves as a cosmological constant.\\
Eigenvalues corresponding to the Jacobian matrix of fixed point E are:
\begin{equation*}
    \Bigg( -2-\lambda^2, \quad -3-\lambda^2, \quad -3-\frac{\lambda^2}{2}, \quad -3-\frac{k\lambda}{2}+(\beta-\lambda)\lambda \Bigg)
\end{equation*}
The point E is of hyperbolic nature. The point exhibits the stable attractor  for $\bigg( (\beta-\lambda)\lambda < (3+\frac{k\lambda}{2}) \bigg)$. However, it exhibits unstable features for $ \bigg( (\beta-\lambda)\lambda > (3+\frac{k\lambda}{2}) \bigg)$.
It has also been observed that the cosmological parameters show nearly steady behaviour at late times with ‘$ln(a)$’ variation. At small redshift ($z$) it is found that the density parameter $\Omega_m$ changes steadily at late times (the reader can also refer \cite{Patil:2022uco,Johnson:2021wou}). This supposition brings us
a practicability to constrain the model in such a way that the term $`\frac{(2\beta - k)}{2} \Omega_m$' in the interaction strength becomes constant, and we termed it as a coupling constant $`$C'. Also, to ease the analysis of the system, we chose particular values of $\beta$ and $k$ where the coupling term $\frac{(2\beta - k)}{2} \Omega_m$ takes values, for example, as C = 0, C = 0.001, C = 0.04. As a result, we observe the dark energy driven accelerated cosmological solution in the three-dimensional time evolution plot of critical point E in Fig. \ref{fig:3D_attractorE}. All the trajectories approaching point E indicate the stable (attractor) behaviour of this point.
%
\par
\begin{figure}[H]
    \centering
    \includegraphics[width=0.8\textwidth]{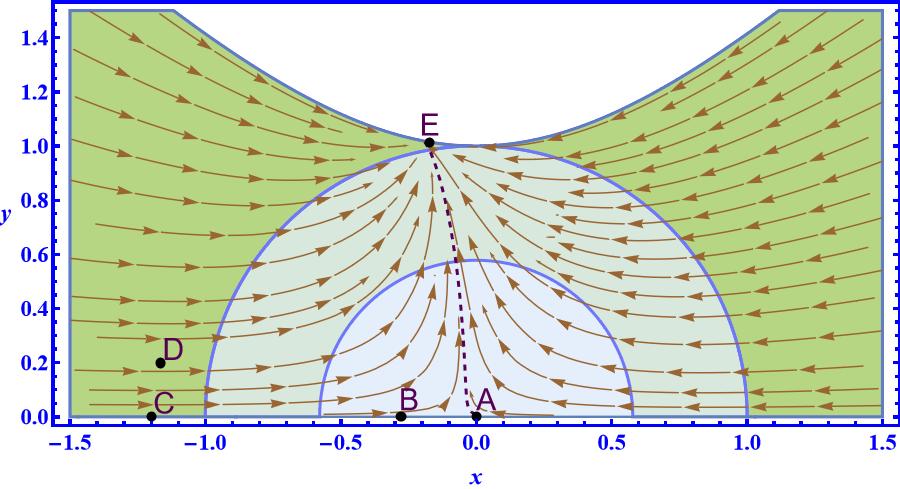}
    \caption{2D phase space portrait of the dynamical system for an equilibrium point E where point E is an attractor for $\lambda=0.4$ and coupling constant $`$C'= 0.001. Points C and D show the saddle behaviour as per the constraints examined before. }
    \label{fig:2dacceleration}
\end{figure}

The scalar field dark energy dominated solution for the critical point E is also represented via 2D phase space portrait in Fig. \ref{fig:2dacceleration}. Here, point E is a phantom-dominated attractor. The purple-dashed line connecting point A with point E features the transition from curvature dominated epoch to the phantom accelerated epoch of the universe. The internal light green region of the phase space portrait of Fig. \ref{fig:2dacceleration} indicates the part where the universe undergoes acceleration with $-1<\omega_{eff}<-\frac{1}{3}$. In contrast, the outer dark green region indicates the phantom-dominated behaviour ($\omega_{eff}<-1$) of point E, where it exhibits a future attractor solution. That is, it is evident from the figure that point E constitutes an everlasting late-time accelerated solution of the universe for the parameter constraint $\lambda^2 >-2$. The equilibrium points B, C, and D act as saddle points in the phase plane. The movement of the trajectories is such that after crossing the neighbourhood of saddle points, they all finally converge towards the attractor point E. In such a case, the equilibrium points A, B, C, and D pose past attractor features in the specified phase space of the system. 

\subsection{Cosmological significance of the model}
\begin{figure}[!ht]
  \centering
  \begin{subfigure}[b]{0.49\linewidth}
    \includegraphics[width=\linewidth]{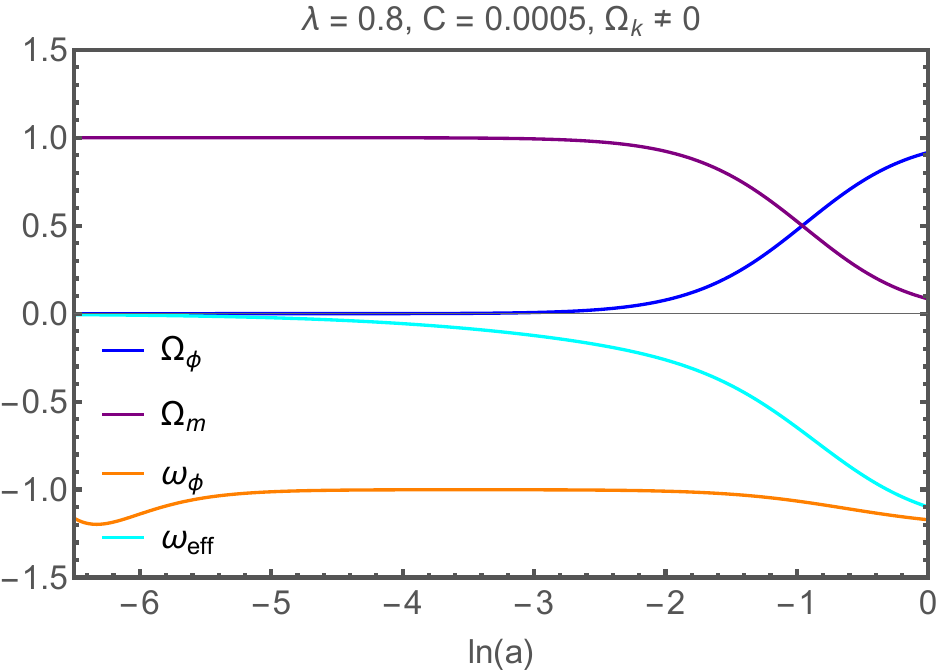}
     \caption{}
     \label{fig:evolution1}
  \end{subfigure}
  \begin{subfigure}[b]{0.49\linewidth}
    \includegraphics[width=\linewidth]{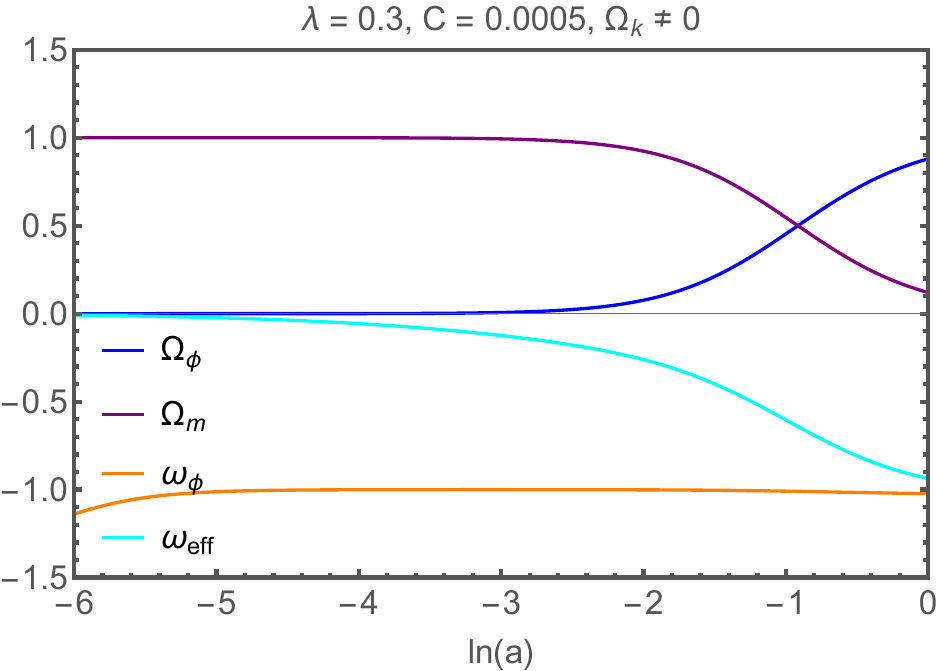}
    \caption{}
    \label{fig:evolution2}
  \end{subfigure}
  \begin{subfigure}[b]{0.5\linewidth}
    \includegraphics[width=\linewidth]{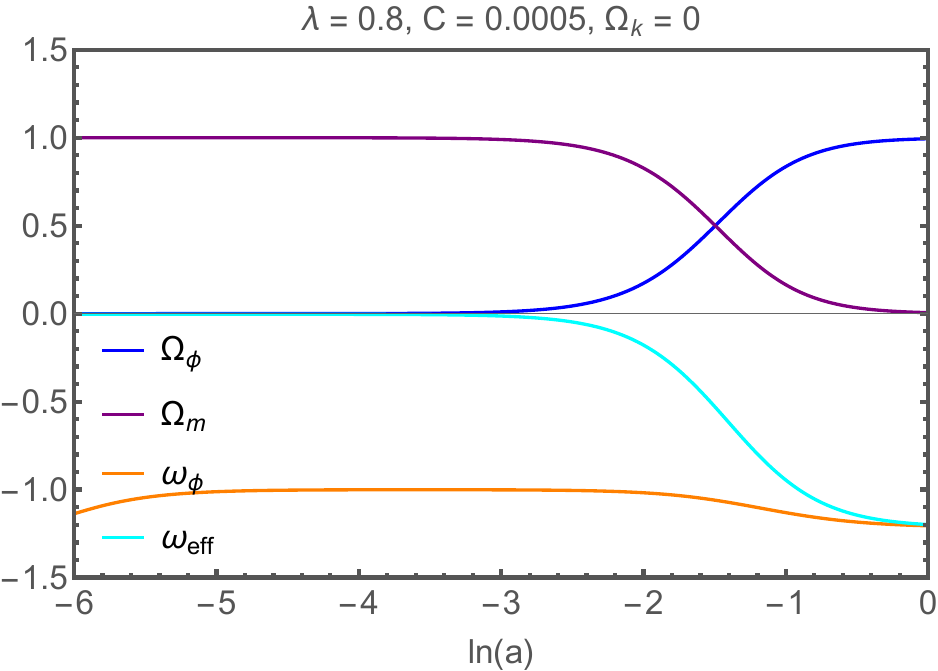}
    \caption{}
    \label{fig:evolution_without curvature}
  \end{subfigure}  
  \caption{Qualitative evolution of cosmological parameters for 4D autonomous system (Eq.\eqref{eqn:autonomous eqns}) for different $\lambda$ values during matter to dark energy transition period.}
    \label{fig:param_evol}
\end{figure}

For any model to be cosmologically significant, the solution of the dynamical system should exhibit a current era of an accelerated expanding universe preceded by a long-enough matter dominated era. To achieve this, we perform the numerical simulations where we choose the following initial conditions:
\begin{equation*}
    x_i= 1.5\times 10^{-4}, \quad y_i= 2.5 \times 10^{-4}, \quad \Omega_{m_i}=0.99, \quad  \Omega_{k_i} = 1+ x_{i}^2 - y_{i}^2 - \Omega_{m_i}.
\end{equation*}
We study the case where $`\lambda$' and coupling constant $`$C' are retained to have positive values. 
We will substantiate the above qualitative analysis by employing the constraints mentioned earlier. In Fig. \ref{fig:param_evol}, we present the qualitative evolution of the fundamental cosmological parameters relative to the 4D autonomous system.
\par
Including the curvature term makes the attributes of the cosmological parameters match with the current observed data at different values of $`\lambda$' and $`$C' compared to the findings with the flat case. In Fig. \ref{fig:param_evol}, for $\lambda =0.8$, $\omega_{eff}$ portrays the transition from quintessence behaviour to phantom field behaviour in the current time. Meaning that it crosses the phantom divide line (Fig. \ref{fig:evolution1}). The above observation justifies the choice of action surveyed in the present study, where the fusion of canonical and non-canonical scalar fields succeeds in achieving crucial observable phenomenons \cite{Hannestad:2002ur,Melchiorri:2002ux,Lima:2003dd,Alam:2004jy,Alam:2003fg,  Wang:2003gz}. On the other hand, for $\lambda=0.3$, $\omega_{eff}$ limits the behaviour to quintessence kind (Fig. \ref{fig:evolution2}).
Furthermore, depending on these choices of parameter values and initial conditions, the DE and DM energy densities at present correspond to the measurements obtained from the various observational analysis \cite{Planck:2015fie,Planck:2015bue,Amendola:2006dg}. We also illustrate the difference between flat and non-flat cases. From Fig. \ref{fig:evolution1} and \ref{fig:evolution_without curvature}, we found that the epoch of dark energy domination starts earlier in the flat coupling case than in the non-flat coupling case. From the observational analysis, the parameter $\Omega_k$ can be constrained \cite{Ryan:2018aif,Yu:2017iju} and one can see its effects on the evolutionary dynamics. Nevertheless, here, we try to show that both scenarios with added coupling can be checked with the data and further analyzed to alleviate the $H_0$ tension \cite{Dinda:2021ffa}. That being the case, we infer that though the presence of non-zero curvature alters the parameters constraints at which the viable cosmological solution is obtained, it does not alter the overall evolution of the universe much. We also notice a matter dominated evolutionary period, long enough for structure formation to occur, accompanied by accelerated expansion driven by dark energy (displaying both; quintessence and phantom character for specific parameter values) at present.
\section{Conclusion}\label{4}
In this work, we extended and examined the dynamics of the DM-DE coupling model developed with the help of the field-fluid analogy method in the case of curved space-time. The fixed point analysis and stability analysis revealed that the equilibrium points of interest are hyperbolic and differ in their stability properties (Table \ref{tab:critical points table}) in curvature space background. Compared to the flat case, including the curvature term affects the cosmological behaviour and produces the results for different values of parameter constraints in the current work.
\par
Point A and point B both show Milne-like solutions. In the case of equilibrium point A, all possible 3D phase portraits display the saddle behaviour (unstable manifold). However, in the case of equilibrium point B, the stability features are shown in Fig. \ref{fig:pointB_stable}. Critical point C also illustrates an accelerated expanding solution under certain constraints on parameters $\beta$ and $k$. The equilibrium point D has equal significance as the point details the late-time accelerated scaling solution that can help to alleviate one of the leading problems in cosmology, i.e., the coincidence problem. It also explains a phantom-dominated and quintessence-dominated accelerating universe under particular constraints on parameters $\beta$, $\lambda$, and $k$. In addition to the above, the curvature part plays its central role when we analyze the equilibrium point E. The curvature alters the eigenvalues of point E and hence, the behaviour and the corresponding analysis. Unlike in the flat case \cite{Patil:2022uco}, where the critical point A delivers a matter-dominated solution, Fig. \ref{fig:2dacceleration} features the transition from curvature dominated point, point A, to the phantom dominated point, point E. Here, all the trajectories' course of action clearly explains the future attractor nature of point E, compelling the other equilibrium points A, B, C, and D to act as past attractors.
\par
Later, we discussed the physical significance of the working model in curved cosmologies via geometrical presentation in Fig. \ref{fig:param_evol}, where we found out that the present-time cosmologically viable universe can be realized in such a scenario as far as we study the background dynamics. However, we have achieved these results for different values of $\beta$, $\lambda$ and $k$ as the effect of curved space-time. In the future, we plan to examine the model's compatibility with the recent observational data in detail.
\section*{Acknowledgment}
This work is partially supported by DST (Govt. of India) Grant No. SERB/PHY/2021057.
The authors thank Mr. Soumya Chakraborty for his thoughtful discussions and critiques. 
\section*{Appendix}\label{Appendix}
Since the eigenvalues of critical points B and D are too complex and challenging to arrange in full size, we struck out a large number of terms. We intended to impart the mannerism in which they emerge during the course of action. The complexity of the eigenvalues indicates their unsuitability to perform the stability analysis of the corresponding fixed points.
\begin{center}
  \textbf{Critical Point B Eigenvalues}:  
\end{center}
\begin{multline*}
    \Bigg[1+\frac{\lambda}{2\beta-k}, -\frac{1}{(12(2\beta-k)^3} \sqrt{(6912 k^7-3456k^9 -497664k^7 \beta^2+... )  +(1152 k^{2} -4608 k \beta +...} ,\\[1.0pt]
    -\frac{1}{(12(2\beta-k)^3} \sqrt{(6912 k^7-96768 k^6 \beta -...  ) +(1152 k^{2} +...)+ (-216 k^2 \beta+432 k \beta^2-...},\\[1.0pt]
    -\frac{1}{(12(2\beta-k)^3} \sqrt{(6912 k^7-3456k^9 -497664k^7 \beta^2+...)   +(-48 k +36 k^3 +96 \beta -...}
    \Bigg].
\end{multline*}
\vspace{0.2cm}
\begin{center}
  \textbf{Critical Point D Eigenvalues}:  
\end{center}
\begin{multline*}
    \Bigg[-\frac{2(18+(2\beta-k -\lambda)(2\beta-k + 2\lambda))}{(2\beta-k + 2\lambda)^2},\\[1.0pt]
    \frac{-1}{(2\beta-k+2\lambda)^3} \sqrt{( 248832k^5 -27869184 k^2 \beta^5+... )  +(4320 k^{4} + 288 k^6 - 228096 k \beta +...)} ,\\[1.0pt]
    \frac{-1}{(2\beta-k+2\lambda)^3} \sqrt{(41472 k^7 -288 k^{10} \lambda -... )  +(57024 k^2 + 55296 k \beta^5 \lambda^2 +...)} ,\\[1.0pt]
    \frac{-1}{(2\beta-k+2\lambda)^3} \sqrt{(2322432 k^6 \beta^2 \lambda -5308416 \beta^7+... )  +(432 k + 24 k^3 + 432 \beta^2 \lambda +...)} ,\\[1.0pt]
    \Bigg].
\end{multline*}
\section*{Data availability statement}
No data associated in the manuscript.

\begin{center}
 \rule{4in}{0.5pt}\\
 \vspace{-13.5pt}\rule{3in}{0.2pt}\\
 \vspace{-13.5pt}\rule{2in}{0.2pt}\\
\end{center}


\begin{spacing}{0.0}
\bibliography{curvaturemain}

\begin{thebibliography}{10}

\bibitem{Riess_1998}
A.~G. Riess, A.~V. Filippenko, and et~al., ``Observational evidence from
  supernovae for an accelerating universe and a cosmological constant,'' {\em
  The Astronomical Journal}, vol.~116, pp.~1009--1038, sep 1998.

\bibitem{Dunkley_2009}
J.~Dunkley, E.~Komatsu, M.~R. Nolta, and et~al., ``{FIVE}-{YEAR} {WILKINSON}
  {MICROWAVE} {ANISOTROPY} {PROBE} {OBSERVATIONS}: {LIKELIHOODS} {AND}
  {PARAMETERS} {FROM} {THE} {WMAP} {DATA},'' {\em The Astrophysical Journal
  Supplement Series}, vol.~180, pp.~306--329, feb 2009.

\bibitem{Boughn:2004zm}
S.~P. Boughn and R.~G. Crittenden, ``{A Detection of the integrated Sachs-Wolfe
  effect},'' {\em New Astron. Rev.}, vol.~49, pp.~75--78, 2005.

\bibitem{PhysRevD.69.103501}
M.~Tegmark, M.~A. Strauss, and et~al., ``Cosmological parameters from sdss and
  wmap,'' {\em Phys. Rev. D}, vol.~69, p.~103501, May 2004.

\bibitem{2007}
W.~J. Percival, S.~Cole, D.~J. Eisenstein, and et~al., ``Measuring the baryon
  acoustic oscillation scale using the sloan digital sky survey and 2df galaxy
  redshift survey,'' {\em Monthly Notices of the Royal Astronomical Society},
  vol.~381, p.~1053–1066, Sep 2007.

\bibitem{DiValentino:2019jae}
E.~Di~Valentino, A.~Melchiorri, O.~Mena, and S.~Vagnozzi, ``{Nonminimal dark
  sector physics and cosmological tensions},'' {\em Phys. Rev. D}, vol.~101,
  no.~6, p.~063502, 2020.

\bibitem{2006}
E.~J. COPELAND, M.~SAMI, and S.~TSUJIKAWA, ``Dynamics of dark energy,'' {\em
  International Journal of Modern Physics D}, vol.~15, p.~1753–1935, Nov
  2006.

\bibitem{article}
L.~Amendola and S.~Tsujikawa, ``Dark energy: Theory and observations,'' {\em
  Dark Energy : Theory and Observations by Luca Amendola and Shinji Tsujikawa.
  Cambridge University Press, 2010. ISBN: 9780521516006}, 01 2010.

\bibitem{Wang_2016}
B.~Wang, E.~Abdalla, F.~Atrio-Barandela, and D.~Pav{\'{o}}n, ``Dark matter and
  dark energy interactions: theoretical challenges, cosmological implications
  and observational signatures,'' {\em Reports on Progress in Physics},
  vol.~79, p.~096901, aug 2016.

\bibitem{Perivolaropoulos:2021jda}
L.~Perivolaropoulos and F.~Skara, ``{Challenges for \ensuremath{\Lambda}CDM: An
  update},'' {\em New Astron. Rev.}, vol.~95, p.~101659, 2022.

\bibitem{Gavela:2009cy}
M.~B. Gavela, D.~Hernandez, L.~Lopez~Honorez, O.~Mena, and S.~Rigolin, ``{Dark
  coupling},'' {\em JCAP}, vol.~07, p.~034, 2009.
\newblock [Erratum: JCAP 05, E01 (2010)].

\bibitem{Yang:2018xah}
W.~Yang, M.~Shahalam, B.~Pal, S.~Pan, and A.~Wang, ``{Constraints on
  quintessence scalar field models using cosmological observations},'' {\em
  Phys. Rev. D}, vol.~100, no.~2, p.~023522, 2019.

\bibitem{Johnson:2021wou}
J.~P. Johnson, A.~Sangwan, and S.~Shankaranarayanan, ``{Observational
  constraints and predictions of the interacting dark sector with field-fluid
  mapping},'' {\em JCAP}, vol.~01, no.~01, p.~024, 2022.

\bibitem{10.1093/mnras/stx2278}
S.~Pan and G.~S. Sharov, ``{A model with interaction of dark components and
  recent observational data},'' {\em Monthly Notices of the Royal Astronomical
  Society}, vol.~472, pp.~4736--4749, 09 2017.

\bibitem{10.1093/mnras/stv1495}
S.~Pan, S.~Bhattacharya, and S.~Chakraborty, ``{An analytic model for
  interacting dark energy and its observational constraints},'' {\em Monthly
  Notices of the Royal Astronomical Society}, vol.~452, pp.~3038--3046, 07
  2015.

\bibitem{salv}
S.~Capozziello, Ruchika, and A.~A. Sen, ``{Model-independent constraints on
  dark energy evolution from low-redshift observations},'' {\em Monthly Notices
  of the Royal Astronomical Society}, vol.~484, pp.~4484--4494, 01 2019.

\bibitem{Olivares:2007rt}
G.~Olivares, F.~Atrio-Barandela, and D.~Pavon, ``{Dynamics of Interacting
  Quintessence Models: Observational Constraints},'' {\em Phys. Rev. D},
  vol.~77, p.~063513, 2008.

\bibitem{PhysRevLett.82.896}
I.~Zlatev, L.~Wang, and P.~J. Steinhardt, ``Quintessence, cosmic coincidence,
  and the cosmological constant,'' {\em Phys. Rev. Lett.}, vol.~82,
  pp.~896--899, Feb 1999.

\bibitem{2020}
O.~Lahav, ``Dark energy: is it “just” einstein’s cosmological constant
  ${\Lambda}$?,'' {\em Contemporary Physics}, vol.~61, p.~132–145, Apr 2020.

\bibitem{RevModPhys.61.1}
S.~Weinberg, ``The cosmological constant problem,'' {\em Rev. Mod. Phys.},
  vol.~61, pp.~1--23, Jan 1989.

\bibitem{Doran:2002ec}
M.~Doran and C.~Wetterich, ``{Quintessence and the cosmological constant},''
  {\em Nucl. Phys. B Proc. Suppl.}, vol.~124, pp.~57--62, 2003.

\bibitem{2009}
S.~d. Campo, R.~Herrera, and D.~Pavón, ``Interacting models may be key to
  solve the cosmic coincidence problem,'' {\em Journal of Cosmology and
  Astroparticle Physics}, vol.~2009, p.~020–020, Jan 2009.

\bibitem{2016}
M.~Bouhmadi-López, J.~Morais, and A.~Zhuk, ``The late universe with non-linear
  interaction in the dark sector: The coincidence problem,'' {\em Physics of
  the Dark Universe}, vol.~14, p.~11–20, Dec 2016.

\bibitem{2014}
H.~E.~S. Velten, R.~F. vom Marttens, and W.~Zimdahl, ``Aspects of the
  cosmological “coincidence problem”,'' {\em The European Physical Journal
  C}, vol.~74, Nov 2014.

\bibitem{Marra:2015iwa}
V.~Marra, ``{Coupling dark energy to dark matter inhomogeneities},'' {\em Phys.
  Dark Univ.}, vol.~13, pp.~25--29, 2016.

\bibitem{Odintsov:2018uaw}
S.~D. Odintsov and V.~K. Oikonomou, ``{Dynamical Systems Perspective of
  Cosmological Finite-time Singularities in $f(R)$ Gravity and Interacting
  Multifluid Cosmology},'' {\em Phys. Rev. D}, vol.~98, no.~2, p.~024013, 2018.

\bibitem{Oikonomou:2019boy}
V.~K. Oikonomou, ``{Classical and loop quantum cosmology phase space of
  interacting dark energy and superfluid dark matter},'' {\em Phys. Rev. D},
  vol.~99, no.~10, p.~104042, 2019.

\bibitem{Yang:2021oxc}
W.~Yang, S.~Pan, L.~Arest\'e~Sal\'o, and J.~de~Haro, ``{Theoretical and
  observational bounds on some interacting vacuum energy scenarios},'' {\em
  Phys. Rev. D}, vol.~103, no.~8, p.~083520, 2021.

\bibitem{Odintsov:2017icc}
S.~D. Odintsov, V.~K. Oikonomou, and P.~V. Tretyakov, ``{Phase space analysis
  of the accelerating multifluid Universe},'' {\em Phys. Rev. D}, vol.~96,
  no.~4, p.~044022, 2017.

\bibitem{Pal:2019bsm}
S.~Pal and S.~Chakraborty, ``{Dynamical system analysis of a three fluid
  cosmological model: an invariant manifold approach},'' {\em Eur. Phys. J. C},
  vol.~79, no.~4, p.~362, 2019.

\bibitem{PhysRevD.104.063517}
D.~Samart, B.~Silasan, and P.~Channuie, ``Cosmological dynamics of interacting
  dark energy and dark matter in viable models of $f(r)$ gravity,'' {\em Phys.
  Rev. D}, vol.~104, p.~063517, Sep 2021.

\bibitem{potting2021coupled}
R.~Potting and P.~M. Sá, ``Coupled quintessence with a generalized interaction
  term,'' 2021.

\bibitem{GonzlezDaz2000CosmologicalMF}
P.~F. Gonz{\'a}lez-D{\'i}az, ``Cosmological models from quintessence,'' {\em
  Physical Review D}, vol.~62, p.~023513, 2000.

\bibitem{refId0}
{Duary, Tanima}, {Dasgupta, Ananda}, and {Banerjee, Narayan}, ``Thawing and
  freezing quintessence models: a thermodynamic consideration,'' {\em Eur.
  Phys. J. C}, vol.~79, no.~11, p.~888, 2019.

\bibitem{PhysRevD.63.103510}
C.~Armendariz-Picon, V.~Mukhanov, and P.~J. Steinhardt, ``Essentials of
  k-essence,'' {\em Phys. Rev. D}, vol.~63, p.~103510, Apr 2001.

\bibitem{Bamba:2012cp}
K.~Bamba, S.~Capozziello, S.~Nojiri, and S.~D. Odintsov, ``{Dark energy
  cosmology: the equivalent description via different theoretical models and
  cosmography tests},'' {\em Astrophys. Space Sci.}, vol.~342, pp.~155--228,
  2012.

\bibitem{Harko2014ArbitrarySA}
T.~Harko, F.~S.~N. Lobo, and M.~K. Mak, ``Arbitrary scalar-field and
  quintessence cosmological models,'' {\em The European Physical Journal C},
  vol.~74, pp.~1--17, 2014.

\bibitem{Oikonomou:2019muq}
V.~K. Oikonomou and N.~Chatzarakis, ``{The Phase Space of $k$-Essence $f(R)$
  Gravity Theory},'' {\em Nucl. Phys. B}, vol.~956, p.~115023, 2020.

\bibitem{ROZASFERNANDEZ2012313}
A.~Rozas-Fernández, ``Kinetic k-essence ghost dark energy model,'' {\em
  Physics Letters B}, vol.~709, no.~4, pp.~313--321, 2012.

\bibitem{Rudra:2014xba}
P.~Rudra, ``{Towards a possible solution for the coincidence problem: f(G)
  gravity as background},'' {\em Int. J. Mod. Phys. D}, vol.~24, no.~02,
  p.~1550013, 2014.

\bibitem{Vinutha:2021jbs}
T.~Vinutha, K.~Sri~Kavya, and K.~Niharika, ``{Bianchi type cosmological models
  in modified theory with exponential functional form},'' {\em Phys. Dark
  Univ.}, vol.~34, p.~100896, 2021.

\bibitem{Clifton:2011jh}
T.~Clifton, P.~G. Ferreira, A.~Padilla, and C.~Skordis, ``{Modified Gravity and
  Cosmology},'' {\em Phys. Rept.}, vol.~513, pp.~1--189, 2012.

\bibitem{Chervon2013ChiralCM}
S.~V. Chervon, ``Chiral cosmological models: Dark sector fields description,''
  {\em arXiv: General Relativity and Quantum Cosmology}, 2013.

\bibitem{Paliathanasis:2018vru}
A.~Paliathanasis, G.~Leon, and S.~Pan, ``{Exact Solutions in Chiral
  Cosmology},'' {\em Gen. Rel. Grav.}, vol.~51, no.~9, p.~106, 2019.

\bibitem{Paliathanasis_2020}
A.~Paliathanasis, ``Dynamics of chiral cosmology,'' {\em Classical and Quantum
  Gravity}, vol.~37, p.~195014, sep 2020.

\bibitem{Patil:2022uco}
T.~Patil, S.~Panda, M.~Sharma, and Ruchika, ``{Dynamics of interacting scalar
  field model in the realm of chiral cosmology},'' {\em Eur. Phys. J. C},
  vol.~83, no.~2, p.~131, 2023.

\bibitem{PhysRevD.103.123517}
P.~M. S\'a, ``Late-time evolution of the universe within a two-scalar-field
  cosmological model,'' {\em Phys. Rev. D}, vol.~103, p.~123517, Jun 2021.

\bibitem{PhysRevD.103.023510}
J.~P. Johnson and S.~Shankaranarayanan, ``Cosmological perturbations in the
  interacting dark sector: Mapping fields and fluids,'' {\em Phys. Rev. D},
  vol.~103, p.~023510, Jan 2021.

\bibitem{BAHAMONDE20181}
S.~Bahamonde, C.~G. Böhmer, S.~Carloni, E.~J. Copeland, W.~Fang, and
  N.~Tamanini, ``Dynamical systems applied to cosmology: Dark energy and
  modified gravity,'' {\em Physics Reports}, vol.~775-777, pp.~1--122, 2018.
\newblock Dynamical systems applied to cosmology: Dark energy and modified
  gravity.

\bibitem{Boehmer:2008av}
C.~G. Boehmer, G.~Caldera-Cabral, R.~Lazkoz, and R.~Maartens, ``{Dynamics of
  dark energy with a coupling to dark matter},'' {\em Phys. Rev. D}, vol.~78,
  p.~023505, 2008.

\bibitem{Caldera-Cabral:2008yyo}
G.~Caldera-Cabral, R.~Maartens, and L.~A. Urena-Lopez, ``{Dynamics of
  interacting dark energy},'' {\em Phys. Rev. D}, vol.~79, p.~063518, 2009.

\bibitem{Chakraborty:2020vkp}
S.~Chakraborty, S.~Mishra, and S.~Chakraborty, ``{Dynamical system analysis of
  three-form field dark energy model with baryonic matter},'' {\em Eur. Phys.
  J. C}, vol.~80, no.~9, p.~852, 2020.

\bibitem{Chakraborty:2020yfe}
S.~Chakraborty, S.~Mishra, and S.~Chakraborty, ``{A dynamical system analysis
  of cosmic evolution with coupled phantom dark energy with dark matter},''
  {\em Int. J. Mod. Phys. D}, vol.~31, no.~01, p.~2150129, 2022.

\bibitem{2021}
A.~Paliathanasis and G.~Leon, ``Dynamics of a two scalar field cosmological
  model with phantom terms,'' {\em Classical and Quantum Gravity}, vol.~38,
  p.~075013, Mar 2021.

\bibitem{Singh:2019enu}
S.~S. Singh and C.~Sonia, ``{Dynamical system perspective of cosmological
  models minimally coupled with scalar field},'' {\em Adv. High Energy Phys.},
  vol.~2020, p.~1805350, 2020.

\bibitem{Mandal:2021ekc}
G.~Mandal, S.~Chakraborty, S.~Mishra, and S.~K. Biswas, ``{Dynamical analysis
  of interacting non-canonical scalar field model},'' 1 2021.

\bibitem{Chakraborty:2021pkp}
S.~Chakraborty, S.~Mishra, and S.~Chakraborty, ``{Dynamical system analysis of
  self-interacting three-form field cosmological model: stability and
  bifurcation},'' {\em Eur. Phys. J. C}, vol.~81, no.~5, p.~439, 2021.

\bibitem{Lazkoz:2006pa}
R.~Lazkoz and G.~Leon, ``{Quintom cosmologies admitting either tracking or
  phantom attractors},'' {\em Phys. Lett. B}, vol.~638, pp.~303--309, 2006.

\bibitem{Pavlov:2013nra}
A.~Pavlov, S.~Westmoreland, K.~Saaidi, and B.~Ratra, ``{Nonflat time-variable
  dark energy cosmology},'' {\em Phys. Rev. D}, vol.~88, no.~12, p.~123513,
  2013.
\newblock [Addendum: Phys.Rev.D 88, 129902 (2013)].

\bibitem{Tot:2022dpr}
J.~Tot, B.~Yildirim, A.~Coley, and G.~Leon, ``{The dynamics of scalar-field
  Quintom cosmological models},'' {\em Phys. Dark Univ.}, vol.~39, p.~101155,
  2023.

\bibitem{Paliathanasis:2022luh}
A.~Paliathanasis and G.~Leon, ``{Hyperbolic inflationary model with nonzero
  curvature},'' {\em Phys. Lett. B}, vol.~834, p.~137407, 2022.

\bibitem{PhysRevD.68.023509}
S.~M. Carroll, M.~Hoffman, and M.~Trodden, ``Can the dark energy
  equation-of-state parameter w be less than $\ensuremath{-}1?$,'' {\em Phys.
  Rev. D}, vol.~68, p.~023509, Jul 2003.

\bibitem{Hannestad:2002ur}
S.~Hannestad and E.~Mortsell, ``{Probing the dark side: Constraints on the dark
  energy equation of state from CMB, large scale structure and Type Ia
  supernovae},'' {\em Phys. Rev. D}, vol.~66, p.~063508, 2002.

\bibitem{Melchiorri:2002ux}
A.~Melchiorri, L.~Mersini-Houghton, C.~J. Odman, and M.~Trodden, ``{The State
  of the dark energy equation of state},'' {\em Phys. Rev. D}, vol.~68,
  p.~043509, 2003.

\bibitem{Lima:2003dd}
J.~A.~S. Lima, J.~V. Cunha, and J.~S. Alcaniz, ``{Constraining the dark energy
  with galaxy clusters x-ray data},'' {\em Phys. Rev. D}, vol.~68, p.~023510,
  2003.

\bibitem{Alam:2004jy}
U.~Alam, V.~Sahni, and A.~A. Starobinsky, ``{The Case for dynamical dark energy
  revisited},'' {\em JCAP}, vol.~06, p.~008, 2004.

\bibitem{Alam:2003fg}
U.~Alam, V.~Sahni, T.~D. Saini, and A.~A. Starobinsky, ``{Is there supernova
  evidence for dark energy metamorphosis ?},'' {\em Mon. Not. Roy. Astron.
  Soc.}, vol.~354, p.~275, 2004.

\bibitem{Wang:2003gz}
Y.~Wang and P.~Mukherjee, ``{Model - independent constraints on dark energy
  density from flux - averaging analysis of type Ia supernova data},'' {\em
  Astrophys. J.}, vol.~606, pp.~654--663, 2004.

\bibitem{Planck:2015fie}
P.~A.~R. Ade {\em et~al.}, ``{Planck 2015 results. XIII. Cosmological
  parameters},'' {\em Astron. Astrophys.}, vol.~594, p.~A13, 2016.

\bibitem{Planck:2015bue}
P.~A.~R. Ade {\em et~al.}, ``{Planck 2015 results. XIV. Dark energy and
  modified gravity},'' {\em Astron. Astrophys.}, vol.~594, p.~A14, 2016.

\bibitem{Amendola:2006dg}
L.~Amendola, G.~Camargo~Campos, and R.~Rosenfeld, ``{Consequences of dark
  matter-dark energy interaction on cosmological parameters derived from SNIa
  data},'' {\em Phys. Rev. D}, vol.~75, p.~083506, 2007.

\bibitem{Ryan:2018aif}
J.~Ryan, S.~Doshi, and B.~Ratra, ``{Constraints on dark energy dynamics and
  spatial curvature from Hubble parameter and baryon acoustic oscillation
  data},'' {\em Mon. Not. Roy. Astron. Soc.}, vol.~480, no.~1, pp.~759--767,
  2018.

\bibitem{Yu:2017iju}
H.~Yu, B.~Ratra, and F.-Y. Wang, ``{Hubble Parameter and Baryon Acoustic
  Oscillation Measurement Constraints on the Hubble Constant, the Deviation
  from the Spatially Flat \ensuremath{\Lambda}CDM Model, the
  Deceleration\textendash{}Acceleration Transition Redshift, and Spatial
  Curvature},'' {\em Astrophys. J.}, vol.~856, no.~1, p.~3, 2018.

\bibitem{Dinda:2021ffa}
B.~R. Dinda, ``{Cosmic expansion parametrization: Implication for curvature and
  H0 tension},'' {\em Phys. Rev. D}, vol.~105, no.~6, p.~063524, 2022.

\end{thebibliography}
\bibliographystyle{ieeetr}
\end{spacing}


\end{document}